\documentclass[journal]{IEEEtran}
\usepackage{cite}
\usepackage{amsmath,amssymb,amsfonts}
\usepackage{algorithmic}
\usepackage{graphicx}
\usepackage{textcomp}
\usepackage[caption=false]{subfig}
\usepackage{enumerate}

\usepackage{bm}
\makeatletter
\AtBeginDocument{\DeclareMathVersion{bold}
\SetSymbolFont{operators}{bold}{T1}{times}{b}{n}
\SetMathAlphabet{\mathrm}{bold}{T1}{times}{b}{n}
\SetMathAlphabet{\mathit}{bold}{T1}{times}{b}{it}
\SetMathAlphabet{\mathbf}{bold}{T1}{times}{b}{n}
\SetMathAlphabet{\mathtt}{bold}{OT1}{pcr}{b}{n}
\SetSymbolFont{symbols}{bold}{OMS}{cmsy}{b}{n}
\renewcommand\boldmath{\@nomath\boldmath\mathversion{bold}}}
\makeatother

\def\BibTeX{{\rm B\kern-.05em{\sc i\kern-.025em b}\kern-.08em
    T\kern-.1667em\lower.7ex\hbox{E}\kern-.125emX}}
    
\DeclareMathOperator*{\diag}{diag}

\begin{document}

\title{A Sequential Game Framework for Target Tracking}

%

\author{\IEEEauthorblockN{1\textsuperscript{st} Daniel Leal}\\
\IEEEauthorblockA{  \textit{School of Computer Science, Faculty of Engineering} \\ 
\textit{University of Sydney, Sydney, Australia}\\
 }
\and 
 \IEEEauthorblockN{2\textsuperscript{nd} Ngoc Hung Nguyen}\\
\IEEEauthorblockA{\textit{Defence Science and Technology (DST) Group}\\ 
\textit{Adelaide, SA 5111, Australia}\\
}
\and 
 \IEEEauthorblockN{3\textsuperscript{rd} Alex Skvortsov}\\
\IEEEauthorblockA{\textit{Defence Science and Technology (DST) Group}\\ 
\textit{Melbourne, VIC 3207, Australia}\\
}
\and 
 \IEEEauthorblockN{4\textsuperscript{th} Sanjeev Arulampalam}\\
\IEEEauthorblockA{\textit{Defence Science and Technology (DST) Group}\\ 
\textit{Adelaide, SA 5111, Australia}\\
}
\and 
\IEEEauthorblockN{5\textsuperscript{th} Mahendra Piraveenan*} \\
\IEEEauthorblockA{\textit{School of Computer Science, Faculty of Engineering}\\ 
\textit{University of Sydney, Sydney, Australia}\\
\textit{Email: Mahendrarajah.Piraveenan@sydney.edu.au}\\
0000-0001-6550-5358}
}

\maketitle

\begin{abstract}
This paper investigates the application of game-theoretic principles combined with advanced Kalman filtering techniques to enhance maritime target tracking systems. Specifically, the paper presents a two-player, imperfect information, non-cooperative,  sequential game framework for optimal decision making for a tracker and an evader. The paper also investigates the effectiveness of this game-theoretic decision making framework by comparing it with single-objective optimisation methods based on minimising tracking uncertainty.  The paper assumes that both the tracker and the evader are intelligent operators capable of complex decision making, and use passive sensors to get noisy bearing information about each other.  Rather than modelling a zero-sum game between the tracker and the evader, which presupposes the availability of perfect information, and therefore unrealistic, in this paper we model both the tracker and  the evader as playing separate zero-sum games at each time step with an internal (and imperfect) model of the other player, which each player constructs based on the imperfect state information they have about the other player. The study defines complex multi-faceted winning criteria for both tracker and evader, and computes winning percentages for both by simulating their interaction for a range of speed ratios.  The results indicate that game theoretic decision making improves the win percentage of the tracker compared to traditional covariance minimization  procedures in all cases, regardless of the speed ratios and the actions of the evader. In the case of the evader, we find that a simpler linear escape action is most effective for the evader  in most scenarios, though the effectiveness of game-theoretic decision making improves when the evader has a more significant speed disadvantage. Overall, the results indicate that the presented sequential-game based decision making framework significantly improves win percentages for a player in scenarios where that player does not have inherent advantages in terms of starting position, speed ratio, or available time (to track / escape), highlighting that game theoretic decision making is particularly useful in scenarios where winning by using more traditional decision making procedures is highly unlikely.  
\end{abstract}


\section{Introduction}\label{introduction}

The ability to efficiently track and predict the movements of elusive targets is a key performance characteristic of many observation systems. Well-known examples include defense and national security systems, space target tracking, and systems for ecological and environmental monitoring (e.g., animal tracking, intruder detection) \cite{ho2022game,ristic2023sensor, yang2020experimental,ristic2022solving,perera2012maritime,li2021auv,gennari2015game,benichou2011intermittent,reddy2021sector}. One of the main technical challenges in developing these systems is the intermittent and unpredictable behavior of the targets, which actively alter their paths to avoid detection \cite{perera2012maritime}.

In maritime defense systems, the capability to efficiently track and predict the movements of target vessels is critical due to the complex nature of maritime environments, and these challenges are compounded by evasive behaviors of targets  that actively alter their paths to avoid detection \cite{perera2012maritime}. Traditional tracking methods, which rely on deterministic models and static assumptions, often struggle to dynamically adapt to intelligent and intermittent maneuvers of such targets (evaders) in an uncertain environment \cite{ho2022game,ristic2004beyond,ristic2023sensor,yang2020experimental,ristic2022solving,perera2012maritime,li2021auv,gennari2015game,springer2022}. This necessitates the exploration of advanced decision-making frameworks that can effectively incorporate strategic interactions between trackers and targets \cite{ho2022game}, motivating the development of an advanced framework that integrates two-sided intelligent decision-making into the target tracking processes.

In this paper, we present a sequential game framework for target tracking, within the context of incomplete information and intelligent, evasive targets.  More specifically, viewing the tracking problem as an information-driven interaction between a tracker and an evader, we integrate the principles of sequential game theory with advanced estimation techniques. Modelling the tracking problem as a pair of inter-dependent zero sum games to be played at each time step, we simulate the behaviour of both tracker and evader on the premise that they both attempt to optimise their outcome against their own imperfect internal models of the other, which they construct based on the incomplete information they have about the other party. This is arguably a more realistic assumption compared to modelling the interaction as a single zero-sum game with complete information~\cite{ristic2022solving}.    We analyse the effectiveness of this game theoretic framework, by comparing game theoretic decision making based on this framework with more traditional, single-objective optimisation-based decision making. The results indicate that the presented sequential game framework is particularly useful for the tracker to improve its win percentages, and overall this framework is more useful for either the tracker or the evader when they have a significant speed disadvantage compared to the other.

The rest of the paper is structured as follows. Section \ref{background} provides a background on bearings-only tracking and game theory, both relevant to this study.  In section \ref{methodology} we describe our methodology, which is divided into six sub-sections: subsection \ref{problem_formulation} defines and formulates the problem to be addressed, while subsection \ref{CKF} provides formal definitions of the Cubature Kalman Filters used,  subsection \ref{game_solution} formulates the game theoretic solution, subsection \ref{strategies} describes the possible actions (decision-making procedures) available to the tracker and evader,  subsection \ref{simulation} describes the simulations undertaken, and subsection \ref{win_criteria} defines and describes the win criteria for the tracker and the evader.  Section \ref{results}  describes the numerical simulations undertaken and presents comparisons of various decision making procedures and related win percentages for both tracker and evader for a range of scenarios, and interprets and discusses the findings. Finally,  section \ref{conclusions} concludes the paper with a summary of findings and suggestions for future research directions.

\section{Background} \label{background}

This paper explores the effectiveness of game-theoretic decision-making in  target tracking  by undertaking  two-dimensional simulations of a tracker and an evader where the tracker tracks an intelligent evading target. The results aim to contribute to the existing body of knowledge by demonstrating the potential of game theory to enhance the effectiveness of target tracking systems.

\subsection{Bearings-Only Tracking}

Bearings-only tracking~\cite{farina1999target, bingol2011bearings} is a challenging nonlinear estimation problem due to the nonlinear relationship between bearing angle measurements and target dynamics (target position and velocity). Various Kalman filtering algorithms~\cite{welch1995introduction,bishop2001introduction,simon2010kalman} have been applied in the literature for bearings-only tracking, including the extended Kalman filter (EKF) \cite{aidala1979kalman}, the modified polar coordinate EKF (MPEKF) \cite{aidala1983utilization, jawahar2016modified}, sigma-point Kalman filters such as the unscented Kalman filter (UKF) and the cubature Kalman filter (CKF) \cite{toloei2014state, wang2009unscented, yang2015bearings}, the pseudolinear and instrumental-variable based Kalman filter~\cite{Nguyen5,Nguyen6}. The particle filter was also considered for bearings-only target tracking (see, e.g.,~\cite{ristic2004beyond}). Each of these algorithms has advantages and disadvantages. Considering the trade-off between the estimation performance and computational efficiency, we select the CKF algorithm for this work.

\subsection{Game Theory}

Game theory, the study of strategic decision-making, was initially developed as a branch of microeconomics \cite{von1944game, osborne2004introduction, kasthurirathna2014topological}. However, it has since been adopted across diverse fields such as evolutionary biology, sociology, psychology, political science, and computer science \cite{rasmusen1994games, kasthurirathna2015emergence, kasthurirathna2013evolution, thedchanamoorthy2014influence, kasthurirathna2014influence, uddin2016set}. Game theory is applied to understand various phenomena and behavioral patterns in human societies and socio-economic systems, including the emergence and maintenance of cooperation in communities and organizations \cite{perc2017statistical, bendor1995types, perc2016phase, chen2014optimal, kasthurirathna2015emergence}, the modeling of unethical or criminal behavior \cite{capraro2019evolution, helbing2015saving}, and the decision-making processes involved in vaccination against epidemics \cite{wang2016statistical, chang2020game, chang2019effects}. The broad applicability of game theory stems from the prevalence of strategic decision-making scenarios across these disciplines. Typically, a game in game theory involves two or more players, each with a set of strategies and associated payoffs (or utility values), often represented in a payoff matrix in two-player games. Game theory is divided into two broad domains: non-cooperative and cooperative game theory.

\subsubsection{Nash Equilibrium}

Nash equilibrium is a core concept in non-cooperative game theory. It represents a set of strategies in a strategic game where no player has an incentive to unilaterally deviate, given the payoffs \cite{nash1950equilibrium, kasthurirathna2014optimisation}. Nash equilibrium can exist in both pure and mixed strategies, and a game with a finite number of players and strategies is guaranteed to have at least one Nash equilibrium \cite{nash1950equilibrium}.

Formally, let $(S, f)$ be a game with $n$ players, where $S_i$ is the strategy set for player $i$. The strategy profile $S$ is $S = S_1 \times S_2 \times \dots \times S_n$. Let $f(x) = (f_1(x), \dots, f_n(x))$ be the pay-off function for the strategy set $x \in S$. Suppose $x_i$ is the strategy of player $i$ and $x_{-i}$ is the strategy set of all other players. The strategy set $x^* \in S$ is a Nash equilibrium if no unilateral deviation by any player results in a higher utility for that player \cite{kasthurirathna2015evolutionary}. Formally, $x^*$ is a Nash equilibrium if and only if:

\[
\forall i, x_i \in S_i : f_i(x_i^*, x_{-i}^*) \geq f_i(x_i, x_{-i}^*)
\]

\subsubsection{Zero-Sum Games}

Zero-sum games are a class of non-cooperative games where the total payoffs for all players sum to zero. In two-player games, one player's loss is precisely equal to the other player's gain, and such games are often represented by a payoff matrix showing only one player's outcomes. Zero-sum games can be solved using the mini-max theorem \cite{nash1953two}, which identifies a strategy set that minimizes the maximum possible loss (or maximizes the minimum possible payoff) for each player, sometimes referred to as a "pure saddle point." The stock market, for instance, is often considered a zero-sum game, whereas most economic transactions are non-zero-sum, as both parties perceive a gain in value from the transaction.

\subsubsection{Simultaneous and Sequential Games}

A simultaneous game is either a normal-form or extensive-form game where all players make decisions simultaneously in each iteration, without knowing the other players' choices \cite{binmore2007playing, hart1992games}. In contrast, a sequential game is an extensive-form game where players choose their strategies in a predefined order. For example, a negotiation process might be modeled as a sequential game if one party has the privilege of making the first offer, with other parties responding subsequently. In sequential games, some players can observe the actions of others before making decisions. If a player can observe every move of all previous players, the game has "perfect information"; otherwise, it is a game of "imperfect information." Sequential games are often used to model bargaining or negotiation processes.

\subsubsection{Subgames}

A subgame is a subset of a sequential game in which every player knows all previous actions of others \cite{osborne2004introduction}. In other words, a subgame is a portion of the game tree where the initial decision node has perfect information.

\section{Methodology}  \label{methodology}

\subsection{Problem Formulation} \label{problem_formulation}

This study addresses the challenge of target tracking by specifically focusing on scenarios where an intelligent evader actively attempts to avoid detection by a tracker. In this section we explain how we computationally and  mathematically  formulate the problem.

We consider a two-dimensional tracking scenario with a tracker and an intelligent evader. The position and velocity of the tracker at discrete-time~$k$ are denoted by~$\mathbf{p}_{T,k}=[x_{T,k},y_{T,k}]^T$ and $\mathbf{v}_{T,k}=v_T[\cos(\theta_{T,k}), \sin(\theta_{T,k})]^T$, where $v_T$ is the speed of the tracker (which is assumed to be constant) and $\theta_{T,k}$ is the heading angle of the tracker at time~$k$. The state of the tracker at time~$k$ is denoted as $\mathbf{x}_{T,k}=[\mathbf{p}_{T,k}^T, \mathbf{v}_{T,k}^T]^T$. 
Similarly, the position and velocity of the evader at time~$k$ are denoted by~$\mathbf{p}_{E,k}=[x_{E,k},y_{E,k}]^T$ and $\mathbf{v}_{E,k}=v_E[\cos(\theta_{E,k}), \sin(\theta_{E,k})]^T$, where $v_E$ is the speed of the evader (which is assumed to be constant) and $\theta_{E,k}$ is the heading angle of the evader at time~$k$. The state of the evader at time~$k$ is denoted as $\mathbf{x}_{E,k}=[\mathbf{p}_{E,k}^T, \mathbf{v}_{E,k}^T]^T$. The tracker and evader can commit to one  of a  possible number of actions when they become aware of the other for the first time. Their possible actions are described in subsection \ref{strategies}.

For simplicity, the motions of both the tracker and evader are discretised, and the tracker and evader are assumed to move from time~$k$ to $k+1$ along a straight line at heading angles~$\theta_{T,k}$ and~$\theta_{E,k}$ respectively. Therefore, we have
\begin{equation}
\mathbf{p}_{T,k+1}=\mathbf{p}_{T,k}+\Delta\,\mathbf{v}_{T,k} 
\end{equation}
\begin{equation}
\mathbf{p}_{E,k+1}=\mathbf{p}_{E,k}+\Delta\,\mathbf{v}_{E,k},
\end{equation}
where~$\Delta$ denotes the sampling interval.

At time~$k$, the tracker collects a noisy bearing measurement about the evader which is modeled as
\begin{equation}
\tilde{z}_{E,k}=\arctan\frac{y_{E,k}-y_{T,k}}{x_{E,k}-x_{T,k}}+n_{E,k}
\label{eq:bearingMeasureOnEvader}
\end{equation}
where~$n_{E,k}$ is the measurement noise  which is assumed to be an independent zero-mean Gaussian variable with standard derivation~$\sigma_E$.

On the other hand, the evader also collects a noisy bearing measurement about the tracker at time~$k$
\begin{equation}
\tilde{z}_{T,k}=\arctan\frac{y_{T,k}-y_{E,k}}{x_{T,k}-x_{E,k}}+n_{T,k}
\label{eq:bearingMeasureOnTracker}
\end{equation}
where~$n_{T,k}$ is the measurement noise. Here,~$n_{T,k}$ is assumed to be an independent zero-mean Gaussian variable with standard derivation~$\sigma_T$.

The tracker utilises the CKF algorithm to estimate the evader's state based on its collected noisy bearing measurements about the evader. The evader, conversely,  also utilises the CKF algorithm to estimate the tracker's state based on its collected noisy bearing measurements about the tracker. In addition, the evader maintains an internal model of the tracker by simulating the tracker using its own information about the tracker in order to mimic the zero-sum game that is played on the actual tracker.

\subsection{State estimation based on the application of CKF} \label{CKF}
This section summaries the CKF algorithm utilised by the tracker to estimate the evader's state using its collected noisy bearing measurements about the evader. Note that the CKF is also used by the evader to estimate the tracker's state in the same manner by switching the subscript~$T$ (notating the tracker) and~$E$ (notating the evader).

The CKF operates in two main steps: prediction and update. At the beginning of time~$k$, the tracker has the information about the state estimate~$\hat{\mathbf{x}}_{E,k-1|k-1}=[\hat{\mathbf{p}}_{E,k-1|k-1}^T \hat{\mathbf{v}}_{E,k-1|k-1}^T]^T$and covariance~$\mathbf{P}_{E,k-1|k-1}$ of the evader at the previous time~$k-1$ as well as the heading angle~$\theta^{*}_{E,k-1}$ of a simulated evader (i.e, an internal model of the evader at the tracker) which resulted from playing a zero-sum game (refer to Section~\ref{strategies} for details). Using this information, we can predict the state at time~$k$ as
\begin{equation}
\hat{\mathbf{x}}_{E,k|k-1}=\hat{\mathbf{x}}^{*}_{E,k}=\begin{bmatrix}
\mathbf{p}^{*}_{E,k}\\
\mathbf{v}^{*}_{E,k}
\end{bmatrix}\label{eq:CKF:Predict-a}\\
\end{equation}
with
\begin{subequations}
\begin{align}
\mathbf{p}^{*}_{E,k}&=\hat{\mathbf{p}}_{E,k-1|k-1}\nonumber\\
&\quad+\|\hat{\mathbf{v}}_{E,k-1|k-1}\|\Delta[\cos(\theta^{*}_{E,k-1}), \sin(\theta^{*}_{E,k-1})]^T,\\
\mathbf{v}^{*}_{E,k}&=\|\hat{\mathbf{v}}_{E,k-1|k-1}\|[\cos(\theta^{*}_{E,k-1}), \sin(\theta^{*}_{E,k-1})]^T,
\end{align}
\end{subequations}
and the predicted covariance is approximated by
\begin{equation}
\mathbf{P}_{E,k|k-1}=\mathbf{F}\mathbf{P}_{E,k-1|k-1}\mathbf{F}^T.\label{eq:CKF:Predict-b}
\end{equation}
Note that~$\mathbf{F}$ is the state transition matrix defined as 
\begin{equation}
\mathbf{F}=\begin{bmatrix}
\mathbf{I}_{2\times2} & \Delta\mathbf{I}_{2\times2}\\
\mathbf{0}_{2\times2} & \mathbf{I}_{2\times2}
\end{bmatrix}
\end{equation}
where $\mathbf{I}$ and $\mathbf{0}$ are identity and zero matrices with the subscript being their dimensions. 

The CKF update step involves calculations of the cubature points, predicted measurement, innovation covariance matrix, cross-covariance matrix, Kalman gain, and updated state estimate and covariance.

First, the cubature points of the predicted density are calculated as

\begin{equation}
\boldsymbol{\alpha}^{(i)}_{E,k|k-1} = \hat{\mathbf{x}}_{E,k|k-1} + 2\sqrt{\mathbf{P}_{E,k|k-1}} \boldsymbol{\xi}^{(i)}
\label{eq:cubature_points}
\end{equation}
for \(i = 1,\dots,8\), where
\begin{equation}
\boldsymbol{\xi}^{(1)} = \begin{bmatrix} 1 \\ 0 \\ 0 \\ 0 \end{bmatrix}, \quad \boldsymbol{\xi}^{(2)} = \begin{bmatrix} 0 \\ 1 \\0 \\0 \end{bmatrix}, \quad \boldsymbol{\xi}^{(3)} = \begin{bmatrix} 0 \\ 0 \\ 1 \\ 0 \end{bmatrix}, \quad \boldsymbol{\xi}^{(4)} = \begin{bmatrix} 0 \\ 0 \\ 0 \\ 1 \end{bmatrix} ,
\label{eq:cubature_vectors}
\end{equation}
and $\boldsymbol{\xi}^{(5)}=-\boldsymbol{\xi}^{(1)}$, $\boldsymbol{\xi}^{(6)}=-\boldsymbol{\xi}^{(2)}$, $\boldsymbol{\xi}^{(7)}=-\boldsymbol{\xi}^{(3)}$, and~$\boldsymbol{\xi}^{(8)}=-\boldsymbol{\xi}^{(4)}$.

The predicted measurement is then computed as:
\begin{equation}
\hat{z}_{E,k|k-1} = \frac{1}{8} \sum_{i=1}^{8} \beta^{(i)}_{E,k|k-1}
\label{eq:predicted_measurement}
\end{equation}
where
\begin{equation}
\beta^{(i)}_{E,k|k-1} = \arctan \left( \frac{[\boldsymbol{\alpha}^{(i)}_{k|k-1}]_2 - y_{T,k}}{[\boldsymbol{\alpha}^{(i)}_{k|k-1}]_1 - x_{T,k}} \right).
\label{eq:beta_calculation}
\end{equation}
Here, $[\cdot]_i$ denotes the $i$-th component of the vector.

The innovation variance and the cross-covariance matrix are given by:
\begin{equation}
P^{zz}_{E,k|k-1} = \frac{1}{8} \sum_{i=1}^{8} (\beta^{(i)}_{E,k|k-1} - \hat{z}_{E,k|k-1})^2 + \sigma_E^2
\label{eq:innovation_variance}
\end{equation}
\begin{equation}
\mathbf{P}^{xz}_{E,k|k-1} = \frac{1}{8} \sum_{i=1}^{8} (\boldsymbol{\alpha}^{(i)}_{E,k|k-1} - \hat{\mathbf{x}}_{E,k|k-1})(\beta^{(i)}_{E,k|k-1} - \hat{z}_{E,k|k-1}).
\label{eq:cross_covariance}
\end{equation}

Finally, the state and covariance update are calculated as
\begin{equation}
\hat{\mathbf{x}}_{E,k|k} = \hat{\mathbf{x}}_{E,k|k-1} + \mathbf{W}_{E,k}(\tilde{z}_{E,k} - \hat{z}_{E,k|k-1})
\label{eq:state_update}
\end{equation}
\begin{equation}
\mathbf{P}_{E,k|k} = \mathbf{P}_{E,k|k-1} - \mathbf{W}_{E,k} P^{zz}_{E,k|k-1} \mathbf{W}_{E,k}^\top,
\label{eq:covariance_update}
\end{equation}
where~$\mathbf{W}_{E,k}$ is the Kalman gain computed as
\begin{equation}
\mathbf{W}_{E,k} = \frac{\mathbf{P}^{xz}_{E,k|k-1}}{P^{zz}_{E,k|k-1}}.
\label{eq:kalman_gain}
\end{equation}

{
\subsection{Game theoretic formulation} \label{game_solution}

As mentioned above, when the tracker and evader become aware of each other at $k=0$, they could take an action which is to engage in a zero-sum game with their own internal model of the other player, and choose their headings accordingly. This is the primary solution proposed in this paper, while the non-game theoretic approaches such as linear escape action, or unilateral optimisation of the covariance matrices are used only as null models to be compared against the game theoretic solutions.

Therefore, let us formulate the zero-sum game to be played at each time step. Formally, the game at each time step \(k\) is defined by the triple \((\mathcal{A}_k, \mathcal{B}_k, \mathbf{U}_k)\), where:

\begin{itemize}
    \item \(\mathcal{A}_k = \{\theta_{k+1}^p(m); m = 1, \ldots, N\}\) is the set of admissible actions for the tracker.
    \item \(\mathcal{B}_k = \{\theta_{k+1}^e(n); n = 1, \ldots, N\}\) is the set of admissible actions for the evader.
    \item \(\mathbf{U}_k\) is the payoff matrix.
\end{itemize}

The tracker (or the evader's internal model of the tracker) aims to solve a min-max optimization problem, while the evader  (or the tracker's internal model of the evader) solves a max-min problem.  The payoff and solution of the game is described in the following subsection, especially in equations \ref{eq27},   \ref{eq28} and \ref{eq18}. The game continues until the estimation accuracy $\psi_e$ reaches a predefined threshold:

\begin{equation} \label{eq2}
\psi_e=\sqrt{\text{trace}(\mathbf{P}_{k|k})} < \eta
\end{equation}

where \(\eta\) is a user-defined threshold.

Once this threshold is reached, the solution of the game is used by the tracker /  evader  to choose their headings for the next time step in the simulation.

\subsection{Possible actions of the tracker and evader} \label{strategies}

We consider and model a number of actions the tracker and evader can undertake, when  they become aware of each other at $k=0$.

For simplicity, we assume that the evader is aware of being tracked at the exact same point of time when the tracker begins tracking it.  Therefore, the evader begins to take evasive action from the time when the tracker begins tracking it, and this is considered the starting point of the simulation,  $k=0$.

A player can undertake a number of possible actions in order to achieve their respective goals: in the case of the tracker, this involves actions to track down or capture the evader, and in the case of the evader, this involves actions to escape from being tracked down or captured.  All these actions can be generically called `strategies'~\cite{schelling1980strategy}. However, the phrase `strategy' has specific connotations in game theory~\cite{osborne2004introduction}.  In this paper we reserve the term `strategy' to mean an action choice that is considered within the game theoretic framework: that is, a particular action which has a corresponding payoff and could  be part of a set of strategies which could result in equilibrium. We do not use the term `strategy' to denote any other choice a tracker or evader might make which is not included as one of the possible actions within the game that is played.

Nevertheless, to quantitatively highlight the benefits of using game theory, we need to have a null model -  which should involve some actions that both the tracker and evader will take if they do not use the game theoretic approach.  Such actions are simply called `actions' in this paper, to be contrasted with `strategies' which are used only within the context of game theory.

We therefore assume that the tracker can take one of the following actions  when it begins tracking an evader (at $k=0$)

\begin{enumerate}[(i)]

    \item Choosing a heading which minimises the covariance trace of the state estimate of the evader at each time step.  The intention here is to minimize the uncertainty about the state estimate of the evader. Note that, in this mode, instead of eq.~\ref{eq:CKF:Predict-a} and eq.~\ref{eq:CKF:Predict-b}, a standard prediction step of the CKF algorithm will be used as
\begin{subequations}
\begin{align}
\hat{\mathbf{x}}_{E,k|k-1}&=\mathbf{F}\hat{\mathbf{x}}_{E,k-1|k-1}\\
\mathbf{P}_{E,k|k-1}&=\mathbf{F}\mathbf{P}_{E,k-1|k-1}\mathbf{F}^T+\mathbf{Q}_E
\end{align}
\end{subequations}
where $\mathbf{Q}_{E}$ is the covariance matrix of the process noise~$\mathbf{w}_{E,k-1}\sim\mathcal{N}(\mathbf{0},\mathbf{Q}_E)$ given by
\begin{equation}
\mathbf{Q}_E=\begin{bmatrix}
\frac{\Delta^3}{3}\mathbf{Q}_{E,\circ} & \frac{\Delta^2}{2}\mathbf{Q}_{E,\circ}\\
\frac{\Delta^2}{2}\mathbf{Q}_{E,\circ} & \Delta\mathbf{Q}_{E,\circ}.
\end{bmatrix}
\end{equation}
Here, $\mathbf{Q}_{E,\circ}=\diag(q_{E,x},q_{E,y})$ with $q_{E,x}$ and $q_{E,y}$ being the power spectral densities of the process noise in $x$- and $y$-coordinates.
    \item Choosing a heading by playing  a zero-sum game with its internal model of the evader, constructed using noisy passive sensor measurements, at each time step.

\end{enumerate}

The second action involves the employment of game theory in decision making, while the first action represents an optimisation process based on the incomplete information available to the tracker, and does not involve game theory.

Similarly, we assume that the evader  can take one of the following actions  when it begins taking evasive actions (at $k=0$)

\begin{enumerate}[(i)]

\item Attempting  a linear escape. This action represents the evader not actively taking evasive action, but continuing with the same heading that it had at $k=0$ with the speed of $v_E$.  If the tracker speed $v_T$ is lower than the evader speed ( $v_T \le v_E$ ), the evader may escape. 

\item Choosing a heading by playing  a zero-sum game with its internal model of the tracker, constructed using noisy passive sensor measurements, at each time step.

\end{enumerate}

These possible actions are described in more detail below.

\vspace{2ex}

\subsubsection{Possible actions of the tracker}

\paragraph*{\textbf{Minimising covariance action of the tracker}}
In this strategy, the tracker chooses a heading angle $\theta_{T,k}$ at each time step $k$ such that it will reduce estimation uncertainty regarding the evader by  minimizing the covariance trace of its position estimate of the evader \(\mathbf{P}^{xy}_{E,k+1|k+1}\) at time $k+1$, where \(\mathbf{P}^{xy}_{E,k+1|k+1}\) is the submatrix of the state covariance matrix \(\mathbf{P}_{E,k+1|k+1}\) defined as
\begin{equation}
\mathbf{P}^{xy}_{E,k+1|k+1} =\begin{bmatrix}
[\mathbf{P}_{E,k+1|k+1}]_{1,1}  & [\mathbf{P}_{E,k+1|k+1}]_{1,2}\\
[\mathbf{P}_{E,k+1|k+1}]_{2,1}  & [\mathbf{P}_{E,k+1|k+1}]_{2,2}
\end{bmatrix}.
\end{equation}

At time~$k$, the evader state estimate $\hat{\mathbf{x}}_{E,k|k}$ and the associated covariance \(\mathbf{P}_{E,k|k}\) is available at the tracker. The tracker then evaluates potential movements by simulating the covariance matrix of the evader state estimate for each hypothesised heading angle $\theta^{(l)}_{T,k}$ as follows.

\begin{itemize}
\item Compute the predicted state and covariance:
\begin{subequations}
\begin{align}
\hat{\mathbf{x}}_{E,k+1|k}&=\mathbf{F}\hat{\mathbf{x}}_{E,k|k}\label{eq:PredictedStateEstimate}\\
\mathbf{P}_{E,k+1|k}&=\mathbf{F}\mathbf{P}_{E,k|k}\mathbf{F}^T+\mathbf{Q}_E.
\end{align}
\label{eq:SimulatedGameStart}
\end{subequations}
\item Compute the Cubature points:
\begin{equation}
\boldsymbol{\alpha}^{(i)}_{E,k+1|k} = \hat{\mathbf{x}}_{E,k+1|k} + 2\sqrt{\mathbf{P}_{E,k+1|k}} \boldsymbol{\xi}^{(i)}
\end{equation}
for $i = 1,\dots,8$.
\item Calculate the hypothesised position of the tracker corresponding to~$\theta^{(l)}_{T,k}$:
\begin{equation}
\mathbf{p}^{(l)}_{T,k+1}=\mathbf{p}_{T,k}+v_T\Delta[\cos(\theta^{(l)}_{T,k}), \sin(\theta^{(l)}_{T,k})]^T.
\end{equation}
\item Compute the predicted measurement using the hypothesised tracker position:
\begin{equation}
\hat{z}^{(l)}_{E,k+1|k} = \frac{1}{8} \sum_{i=1}^{8} \beta^{(l,i)}_{E,k+1|k}
\end{equation}
where
\begin{equation}
\beta^{(l,i)}_{E,k+1|k} = \arctan \left( \frac{[\boldsymbol{\alpha}^{(i)}_{k+1|k}]_2 - [\mathbf{p}^{(l)}_{T,k+1}]_2}{[\boldsymbol{\alpha}^{(i)}_{k|k-1}]_1 - [\mathbf{p}^{(l)}_{T,k+1}]_1}\right).
\end{equation}

\item Compute the innovation variance and the cross-covariance matrix:
\begin{equation}
P^{zz(l)}_{E,k+1|k} = \frac{1}{8} \sum_{i=1}^{8} (\beta^{(l,i)}_{E,k+1|k} - \hat{z}^{(l)}_{E,k+1|k})^2 + \sigma_E^2
\end{equation}
\begin{equation}
\mathbf{P}^{xz(l)}_{E,k+1|k} = \frac{1}{8} \sum_{i=1}^{8} (\boldsymbol{\alpha}^{(i)}_{E,k+1|k} - \hat{\mathbf{x}}_{E,k+1|k})(\beta^{(l,i)}_{E,k+1|k} - \hat{z}^{(l)}_{E,k+1|k}).
\end{equation}
\item Compute the Kalman gain:
\begin{equation}
\mathbf{W}^{(l)}_{E,k+1} = \frac{\mathbf{P}^{xz(l)}_{E,k+1|k}}{P^{zz(l)}_{E,k+1|k}}.
\end{equation}
\item Compute the updated covariance matrix:
\begin{equation}
\mathbf{P}^{(l)}_{E,k+1|k+1} = \mathbf{P}_{E,k+1|k} - \mathbf{W}^{(l)}_{E,k+1} P^{zz(l)}_{E,k+1|k} (\mathbf{W}^{(l)}_{E,k+1})^\top.
\end{equation}
\item Extract the updated covariance matrix of the evader position estimate:
\begin{equation}
\mathbf{P}^{xy(l)}_{E,k+1|k+1} =\begin{bmatrix}
[\mathbf{P}^{(l)}_{E,k+1|k+1}]_{1,1}  & [\mathbf{P}^{(l)}_{E,k+1|k+1}]_{1,2}\\
[\mathbf{P}^{(l)}_{E,k+1|k+1}]_{2,1}  & [\mathbf{P}^{(l)}_{E,k+1|k+1}]_{2,2}
\end{bmatrix}.
\label{eq:SimulatedGameEnd}
\end{equation}

\end{itemize}

The tracker selects the heading angle \(\theta_{T,k}^{\ast}\) that minimizes the trace of the updated covariance matrix of the evader position estimate as:

\begin{equation} \label{eq20}
\theta_{T,k}^{\ast} = \arg \min_{\theta_{T,k}^{(l)} } \left\{\mathrm{trace}\left(\mathbf{P}^{xy(l)}_{E,k+1|k+1} \right) \right\}.
\end{equation}

\paragraph*{\textbf{Choosing a heading by playing  a zero-sum game}} 
In this strategy, the tracker chooses a heading angle $\theta_{T,k}$ at each time step $k$ by playing a zero-sum game with its internal model of the evader.  Given the evader state estimate $\hat{\mathbf{x}}_{E,k|k}=[\hat{\mathbf{p}}_{E,k|k}^T, \hat{\mathbf{v}}_{E,k|k}^T]^T$ and the associated covariance \(\mathbf{P}_{E,k|k}\), the tracker and the simulated model of the evader (denoted as evader$^{\ast}$ for simplicity) will evaluate all possible headings~$\theta^{(l)}_{T,k}\) for the tracker and $\theta^{(m)}_{E,k}$ for the evader$^{\ast}$.  The tracker attempts to minimize the maximum possible trace of the covariance matrix of the evader position estimate, while the evader$^{\ast}$ attempts to maximize the minimum possible trace of the same matrix.

For a given hypothesised pair of action~$\left\{\theta^{(l)}_{T,k},\theta^{(m)}_{E,k}\right\}$, we calculate the updated covariance matrix of the evader position estimate~$\mathbf{P}^{xy(l,m)}_{E,k+1|k+1}$ as in~(\ref{eq:SimulatedGameStart})--(\ref{eq:SimulatedGameEnd}) except that the predicted evader state estimate in~(\ref{eq:PredictedStateEstimate}) is replaced by the hypothesised state of the evader$^{\ast}$ corresponding to~$\theta^{(m)}_{E,k}$ as
\begin{equation}\label{eq:GameStart}
\hat{\mathbf{x}}_{E,k+1|k}=\hat{\mathbf{x}}^{(m)}_{E,k+1}=\begin{bmatrix}
\mathbf{p}^{(m)}_{E,k+1}\\
\mathbf{v}^{(m)}_{E,k+1}
\end{bmatrix}
\end{equation}
where
\begin{subequations}
\begin{align}
\mathbf{p}^{(m)}_{E,k+1}&=\hat{\mathbf{p}}_{E,k|k}+\|\hat{\mathbf{v}}_{E,k|k}\|\Delta[\cos(\theta^{(m)}_{E,k}), \sin(\theta^{(m)}_{E,k})]^T,\\
\mathbf{v}^{(m)}_{E,k+1}&=\|\hat{\mathbf{v}}_{E,k|k}\|[\cos(\theta^{(m)}_{E,k}), \sin(\theta^{(m)}_{E,k})]^T.
\end{align}
\end{subequations}

In this simulated game, the tracker chooses the heading angle~\(\theta_{T,k}^{\ast}\) that minimizes the maximum trace of the updated covariance matrix, while the evader$^{\ast}$ chooses the heading angle \(\theta_{E,k}^{\ast}\) that maximizes the minimum trace of the updated covariance matrix. This can be formulated as:

\begin{equation} \label{eq27}
\theta_{T,k}^{\ast} = \arg \min_{\theta_{T,k}^{(l)}} \max_{\theta_{E,k}^{(m)}} \left\{ \mathrm{trace}\left(\mathbf{P}_{E,k+1|k+1}^{xy(l,m)}\right) \right\}
\end{equation}

\begin{equation} \label{eq28}
\theta_{E,k}^{\ast} = \arg \max_{\theta_{E,k}^{(m)}} \min_{\theta_{T,k}^{(l)}} \left\{ \mathrm{trace} \left( \mathbf{P}_{E,k+1|k+1}^{xy(l,m)}\right) \right\}
\end{equation}

It is important to note that only the heading angle~$\theta_{T,k}^{\ast}$ for the tracker is of consequence and the tracker will take action to change its heading angle to~$\theta_{T,k}^{\ast}$. On the other hand, the heading angle~$\theta_{E,k}^{\ast}$ for the evader$^{\ast}$ is not of any further consequence, as this is not the `real' evader but a simulated model of the evader inside the tracker. 


\subsubsection{Possible actions of the evader}

\paragraph*{\textbf{Linear escape action of the evader}}
The linear escape action is a simple approach where the evader aims to move away from the tracker in a straight line with an unchanged heading and speed $v_E$ from the moment it is aware of being tracked.  This action assumes that the evader continually moves in the direction of its original velocity,  which could be towards a predetermined destination or simply to flee as quickly as possible. This baseline action is considered naive yet realistic, as it may represent an evader's attempt to return to a safe location or evade capture in the simplest manner.  Clearly, we do not consider a corresponding baseline action for the tracker, as a tracker cannot hope to be successful in tracking by moving with constant heading regardless of the evader's actions.

{Mathematically,  this action is represented by updating the evader's position \(\mathbf{e}_t\) at each time step \(t\) using its constant velocity \(\mathbf{v}_e\). The new position \(\mathbf{e}_k\) is computed as:}

\begin{equation} \label{eq17}
\mathbf{e}_{k+1} = \mathbf{e}_k + \mathbf{v}_E \Delta k
\end{equation} where \(\mathbf{e}_k\) is the current position, \(\mathbf{v}_E\) is the evader's velocity, and \(\Delta k\) is the elapsed time per time step. In this simulation, we assume that $\Delta k = 1$.

\paragraph*{\textbf{Choosing a heading by playing  a zero-sum game}} 
In this approach, the evader chooses a heading angle~$\theta_{E,k}$ at each time step~$k$ by playing a zero-sum game by simulating the tracker using its own information about the tracker. We refer to this simulated tracker as tracker$^*$ for simplicity. The tracker$^*$ has its own CKF algorithm to track the evader as described in~Section~$\ref{CKF}$. Note that the position and velocity of the tracker$*$ is the estimated position and velocity of the real tracker computed by the evader's CKF, i.e., $\hat{\mathbf{x}}_{E,k|k}=[\hat{\mathbf{p}}_{E,k|k}^T, \hat{\mathbf{v}}_{E,k|k}^T]^T$, and the bearing measurement about the evader available at the tracker$^*$ at time~$k$ is generated as

\begin{equation} \label{eq18}
\tilde{z}_{E,k}=\arctan\frac{y_{E,k}-[\hat{\mathbf{p}}_{T,k|k}]_2}{x_{E,k}-[\hat{\mathbf{p}}_{T,k|k}]_1}+n_{E,k}.
\end{equation}
where~$n_{E,k}$ is generated from a zero-mean Gaussian distribution with standard deviation~$\sigma_E$.

The tracker$^*$ then plays a zero-sum game similarly to that described in~(\ref{eq:GameStart})-(\ref{eq28}). It is important to note that only the heading angle~$\theta^*_{E,k}$ chosen for the evader by the game is of consequence and the evader will take action to adjust its heading angle to~$\theta^*_{E,k}$. On the other hand, the heading angle~$\theta^*_{T,k}$ chosen for the tracker$^*$ by the game is not of any further consequence because the tracker$^*$ is not the `real' tracker but a simulated model of the tracker inside the evader.

\subsection{Simulation of tracker and evader} \label{simulation}

The study simulates a two-dimensional maritime environment with infinite dimentions. At time \(k=0\), a tracker and an evader are positioned in a grid space.  The simulation evolves in discrete time steps \(k\). At each step, both players must decide their next course of action  without knowing the other's choice.  The simulation completes after a maximal number of time steps ($k_{max}$). In all experiments described below, $k_{max}=400$. The tracker moves with speed $v_T$, while the evader  moves with speed $v_E$.  It is assumed that they become aware of each other at the same time, at $k=0$. 

\noindent \textbf{Initialisation}: At $k=0$, the tracker's position is randomly initialised within a square grid of $40 \times 40$ which is centred upon $(x=400, y=550)$. Similarly, the evader's  position is randomly initialised within a square grid of $40 \times 40$ which is centred upon $(x=1000, y=1000)$.  Thus, on average, the starting distance between evader and tracker is  $600$ units on the x-axis and $450$ units on the y-axis.

\noindent \textbf{Tracker and evader velocities}: the tracker speed  $v_T$ and evader speed $v_E$ are both fixed for a particular simulation run. We simulated a range of tracker velocities from   $v_T = 1$ unit  per time step, to $v_T = 3$ units per time step. Similarly, we simulated a range of evader  velocities from   $v_E = 1$ unit  per time step, to $v_E = 3$ units per time step. Therefore, the range of $\frac{v_E}{v_T}$ was between $0.333$ to $3.0$.  We considered three discrete velocities for the tracker, and three discrete velocities for the evader, as represented by integer values between 1 and 3 per time step in each case, resulting in five distinct values for   $\frac{v_E}{v_T} $. For each of these speed ratios, $200$ simulation runs were completed, each starting from random points within the grid space as described above.

Figures \ref{s1} and  \ref{s2}  show two sample simulation runs, each undertaken for 400 time steps.  In  Fig. \ref{s1}  the evader takes the linear escape action, while the tracker takes the covariance minimisation action (decides its heading at each time step by minimising the covariance of the evader state estimate). How the tracker perceives the evader to move, based on its noisy bearing measurements, is also shown. The tracker and evader velocities are   $v_T =  3$  units per time step, and  $v_E = 1$ units per time step respectively, and the figure reflects time step $k=100$. Here it could be seen that the evader flees using linear escape, but the tracker is on course to intercept it using its superior speed and reasonably accurate bearing measurement. It is apparent that given enough time, the tracker will intercept the evader and thus will `win'.

 By contrast, in  Fig. \ref{s2}, both players decide on their headings at each time step by playing a zero sum game with their internal models of the other player, respectively, based on their noisy bearing measurements.  How the tracker and evader  perceive each other to move, based on its noisy bering measurements, is also shown.  The tracker and evader velocities are  $v_T =  3$  units per time step, and  $v_E = 3$ units per time step respectively: that is, neither has a speed advantage over the other. In such a scenario, only superior decision making could result in a win for the tracker.  The figure reflects time step $k=150$. Here it could be seen that both parties are using their bearing measurements of the other, and engage in game-theoretic decision making. Thus the tracker has no advantage in decision making procedure either. As it could be expected in such a scenario, it is apparent from the figure that the tracker is unlikely to intercept the evader or win in this instance. It is particularly interesting to note that after the initial decisions based on very noisy measurements by both players, for a while the tracker was moving in a trajectory which would intercept the evader eventually, but as the evader bearing measurements became more  accurate, the evader changed courser significantly, forcing the tracker to change course as well. After that point in time, the tracker attempts to move in a trajectory which would intercept the evader, but given equal speeds and equally good  decision making by the evader, the tracker will be unable to win.
 
 \begin{figure}
    \centering
    \includegraphics[scale=0.45]{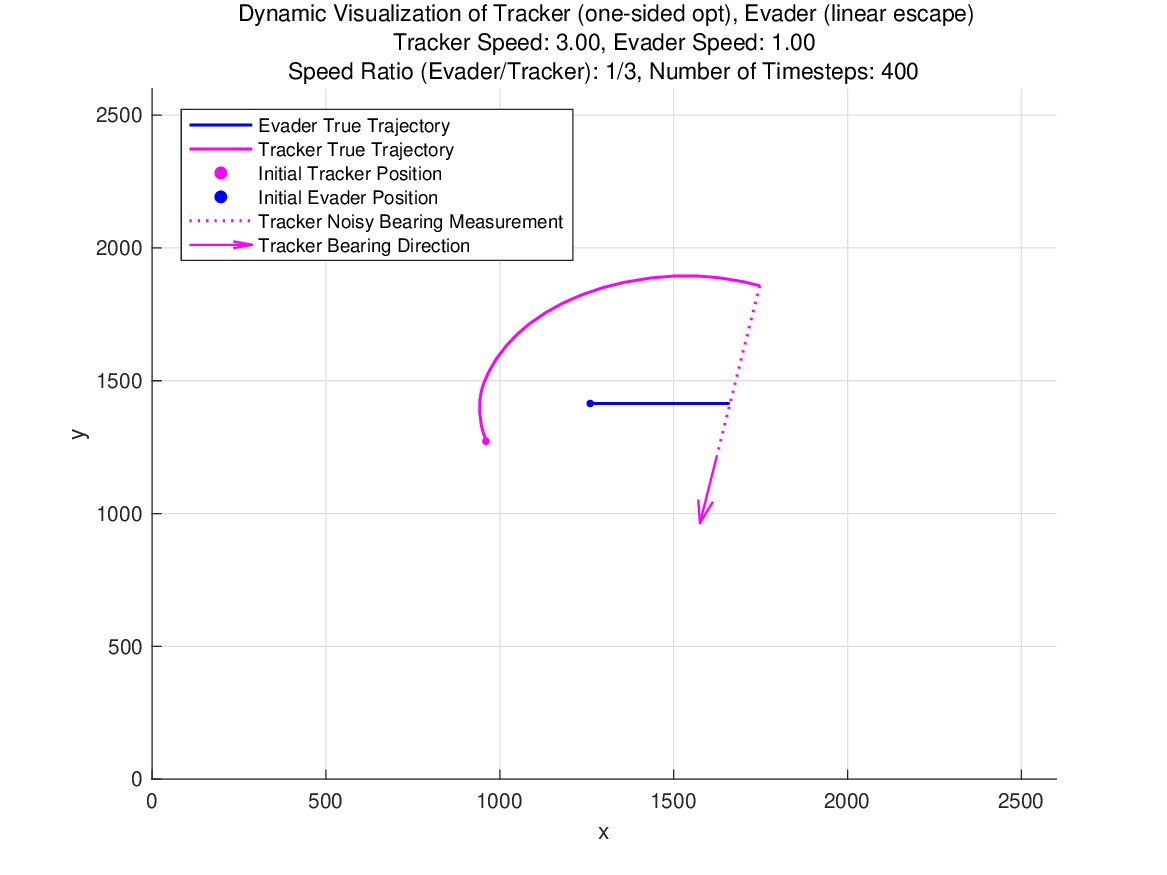}
    \caption{Sample simulation run. Here the tracker undertakes the covariance minimisation action and chooses its heading accordingly at each time step, and the evader undertakes the linear escape action. The tracker speed is $v_T =  3$  units per time step, and the evader speed is $v_E = 1$ units per time step. The figure reflects $k=100$.}
    \label{s1}
\end{figure}

\begin{figure}
    \centering
    \includegraphics[scale=0.45]{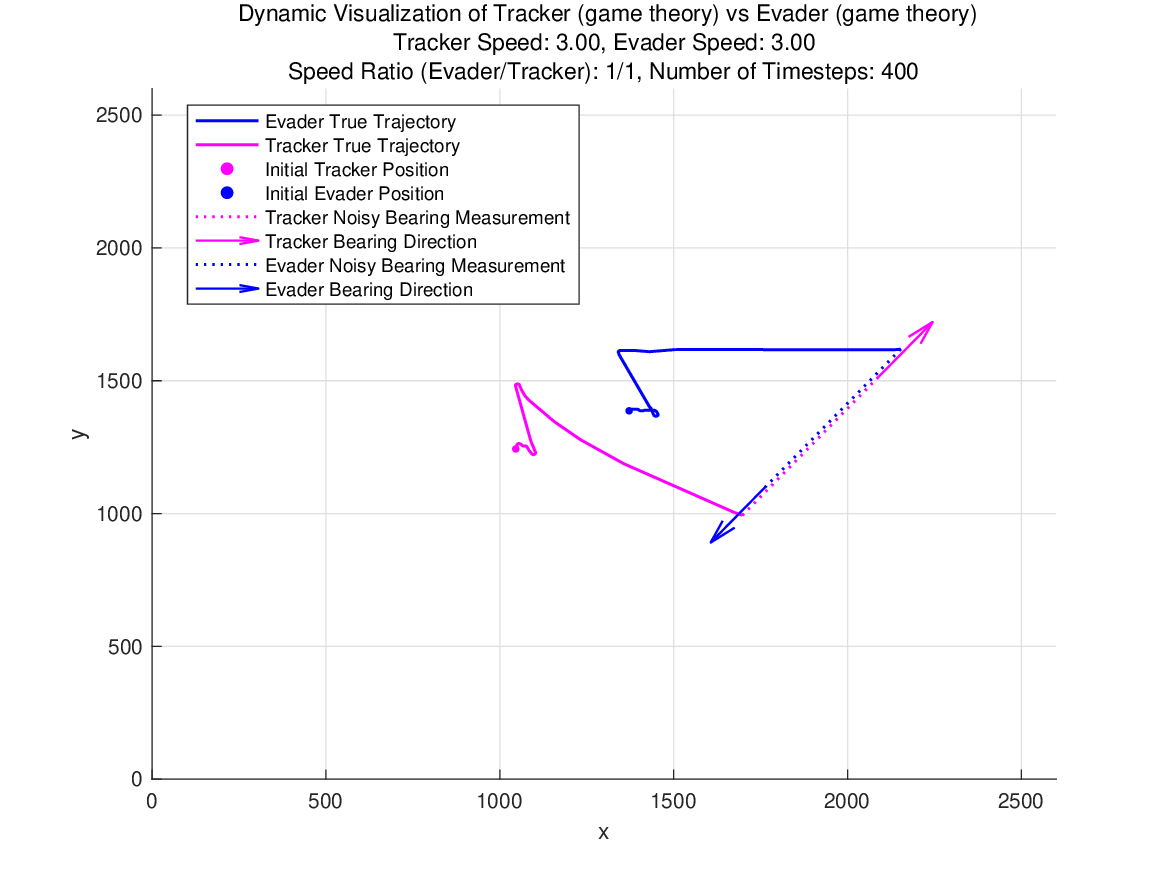}
    \caption{Sample simulation run. Here the tracker and evader both undertake game-theoretic decision making at each time step. The tracker speed is $v_T =  3$   units per time step and the evader speed is $v_E = 3$  units per time step. The figure reflects $k=150$.}
    \label{s2}
\end{figure}

\subsection{Win criteria} \label{win_criteria}

We define and compute a number of metrics to decide which agent  /  player has `won' each round of the simulation.

 \begin{enumerate}
    \item Trace of the covariance matrix of the tracker's  position estimate of the evader
    \item Euclidean distance between the tracker's estimated position of the evader and the actual position of the evader
    \item Euclidean distance between the actual positions of the tracker and the evader. 
  \end{enumerate}

Note well that these metrics are defined from the tracker's perspective. Therefore, a minimal value is favourable to the tracker, while a maximal value is favourable to the evader.

The first metric is the trace of the covariance matrix of the tracker's position estimate of the evader. This metric measures the uncertainty in the tracker's estimated position of the evader.  A lower covariance indicates higher confidence in the tracker's position estimate of the evader, which is crucial for effective tracking. This metric is particularly relevant in scenarios where precise localisation is essential for the tracker to successfully zero in on the evader. By minimising the covariance, the tracker demonstrates its ability to accurately maintain its position estimate of the evader, reducing the uncertainty. Therefore, if the trace of the covariance falls below a predefined minimum threshold, it signifies that the tracker has achieved a high level of certainty in its position estimate, leading to a win for the tracker. Conversely, if it exceeds a maximum threshold, it indicates significant uncertainty, allowing the evader to effectively evade, resulting in a win for the evader.

The second metric is the Euclidean distance between the tracker's estimated position of the evader and the actual position of the evader. This metric reflects the tracker's ability to predict the evader's movements accurately. A smaller distance indicates that the tracker is effectively anticipating the evader's path, a critical factor in successful tracking. This measure directly evaluates the tracker's capability to follow the evader's movements closely, which is the primary goal of the tracker. If this distance falls below a minimum threshold, it suggests that the tracker is successfully predicting the evader's movements, resulting in a win for the tracker. Conversely, if the distance exceeds a maximum threshold, it indicates that the evader has managed to deceive the tracker and maintain a safe distance, resulting in a win for the evader.

The third metric is the Euclidean distance between the actual positions of the tracker and the evader. This metric directly measures the physical proximity between the two agents. A smaller distance indicates that the tracker is close to catching the evader, directly assessing the success of the tracking effort. This criterion is an effective measure of performance because it encapsulates the primary objective of the tracker: to minimise the distance to the evader and to eventually capture it. Conversely, the evader's goal is to maximise this distance and escape. If the distance falls below a minimum threshold, it means the tracker is in close proximity to the evader, signifying a win for the tracker. If the distance exceeds a maximum threshold, it indicates that the evader has successfully maintained a significant distance from the tracker, resulting in a win for the evader.

For each of these metrics, we therefore defined two thresholds;

\begin{itemize}
    \item \textbf{Minimum Threshold}: If the selected metric falls below this threshold, the tracker is declared the winner. This threshold signifies successful tracking or close proximity.
    \item \textbf{Maximum Threshold}: If the selected metric exceeds this threshold, the evader is declared the winner. This threshold indicates that the evader has effectively maintained a safe distance or caused significant uncertainty in the tracker's position estimate.
\end{itemize}

The outcome of  a particular  simulation run is determined based on the following conditions:

\begin{itemize}
    \item \textbf{Tracker Win}: The tracker wins if any of the three metrics fall below its relevant minimum threshold at any point during the simulation.
    \item \textbf{Evader Win}: The evader wins if any of the three metrics exceeds the maximum threshold at any point during the simulation. The evader also wins if the tracker does not win by the end of the simulation (if any of the three metrics did not fall below the minimum threshold at any time during the simulation) because it is assumed that the tracker will stop the tracking effort at the end of the simulation, allowing the evader to escape, which counts as a win for the evader.
\end{itemize}

In addition, the Stop Time  $k_{stop}$ is also recorded at the end of each simulation. Stop Time  $k_{stop}$ is the specific time step at which the simulation stopped. This would be the time at which any of the three metrics mentioned above exceeded either the minimum threshold (tracker win) or maximum threshold (evader win), or it would be $k_{max}$ (evader win).

%
%
%
%
%
%
%
%
%

\section{Numerical Results}  \label{results}

In this section we present the results of the simulation experiments.  The main goal of the simulation experiments was to highlight the set of circumstances under which game-theoretic decision making is particularly useful.  Let us recall from subsection \ref{strategies} that we consider two different actions an evader can undertake at the outset, of which only one involves game theoretic decision making. Similarly we mentioned two different actions that the tracker can undertake, of which only one involves game theoretic decision making. Our aim is to compare the actions which involve game theoretic decision making with those actions which do not involve this, and illustrate the positive difference that game theoretic decision making can bring in.  Given that the evader has two possible actions (linear escape, choosing a heading by playing  a zero-sum game with the simulated tracker) and the tracker has two possible actions (minimising  covariance of the evader state estimation,  choosing a heading by playing  a zero-sum game with the simulated evader) to follow at the outset, four different `action-combinations (action-pairs)' are possible. We look at these combinations one by one.

For each of these action pairs, we undertook simulations for varying tracker and evader velocities.  The tracker speed was simulated as either $v_T = 1$,   $v_T = 2$ or  $v_T = 3$,   with the unit being grid units per time step. Similarly the evader speed was simulated as either $v_E = 1$,   $v_E = 2$ or  $v_E = 3$. This therefore resulted in five speed ratios $R_v  = \frac{v_E}{v_T}$, namely  1/3, 2/3, 1/1, 3/2, and 3/1 (we also tested cases where the velocities were both 2 units per time step, and where both were 3 units per time step. However since these result in a ratio of 1:1 anyhow, we will not present results from those particular speed pairs separately).  For each tracker and evader speed ratio,  we undertook 200 individual simulation runs and averaged our results. For each run we simulated starting points for both tracker and evader randomly within the constraints described in subsection \ref{simulation}.

The results are now presented in terms of  both the tracker's and evader's perspective, and discussed in terms of each possible `action pair' for tracker and evader.

\subsection{Optimal action by tracker}

Here we compare different actions the tracker could take if the evader chooses a particular action.

\subsubsection{Optimal action by tracker if evader chooses linear escape}

Fig. \ref{r_t23e1} plots the percentage of experiments which result in a tracker win (on the y-axis) against different speed ratios (on the x-axis), and compares the two possible actions the tracker could take.  Each datapoint is an average of 200 simulation runs as mentioned earlier. We can first observe that the tracker win percentage is relatively low in all cases: less than 50\% even when the tracker speed is  higher than the evader speed, and the win percentage of tracker decreases further as evader speed increases relatively. Therefore it appears that linear escape is an effective strategy for the evader in general, though if the tracker had more time to track (if the duration of the the simulation  is much higher than 400 time steps), the win percentage would have  favoured the tracker more, at least in cases where the $v_T  \ge v_E$. However it is the relative performance of the tracker in terms of its choice of action that we are primarily interested in.

The figure compares the two possible actions by the tracker, i.e., minimising covariance of the evader state estimate and employing a game theoretic solution. It is apparent that in for all speed ratios, making decisions based on the game theoretic solution (at each time step) helps the tracker get more wins compared to a covariance minimisation strategy.  In particular, when the evader has a strong speed advantage the tracker win percentage could be as low as 10\% if it uses covariance minimisation, but not lower than 20\% if it uses game theoretic decision making. Therefore it is clear that the tracker is hard pressed against an evader who employs the linear escape strategy, and employing a game theoretic solution to decide on its heading helps the tracker perform relatively better, especially when the tracker also has a speed disadvantage.

\begin{figure}
    \centering
    \includegraphics[scale=0.45]{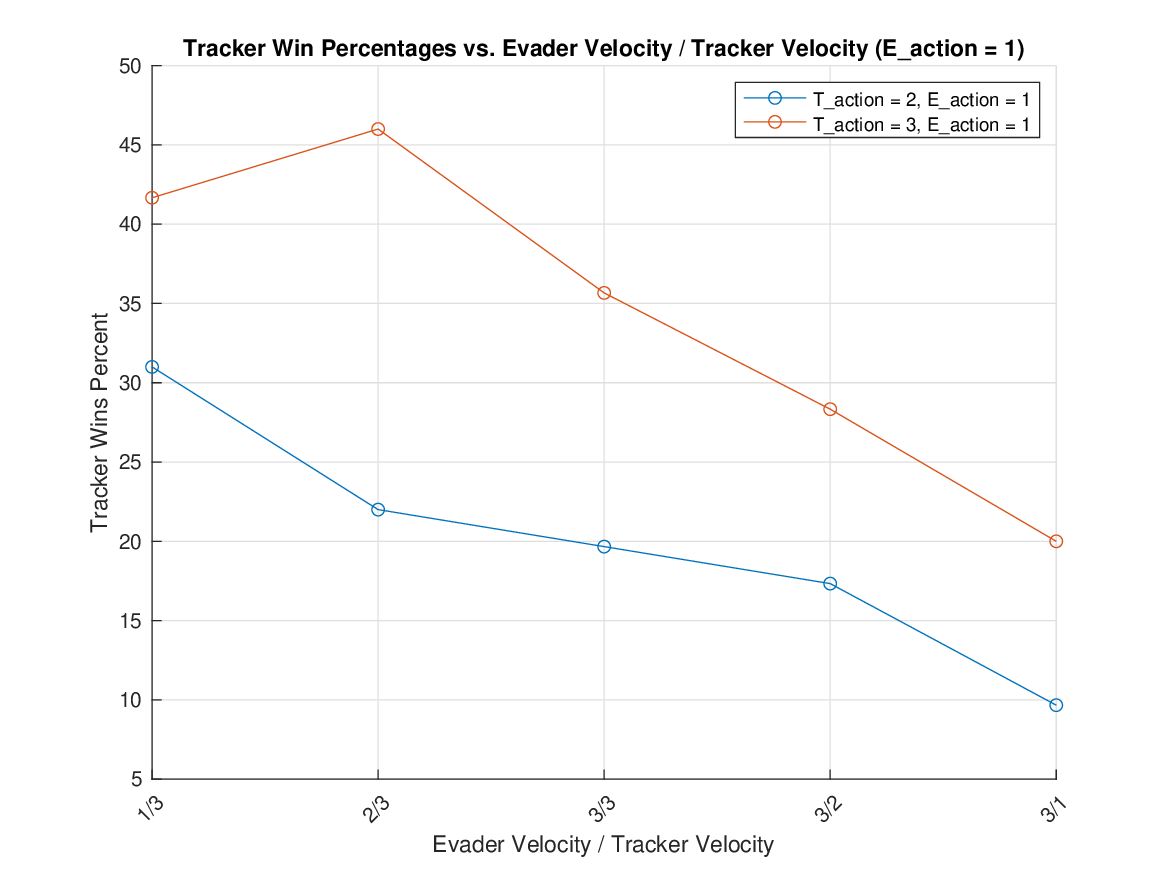}
    \caption{Comparing actions by tracker when evader chooses linear escape. Note that for all speed ratios, the tracker is better off making decisions by using game theory.}
   \label{r_t23e1}
\end{figure}

\subsubsection{Optimal action by tracker if evader undertakes game theoretic decision making}

Now let us look  at which is the better action for the tracker, if the evader were to use game theoretic decision making.  Recall that when the evader commits to the action of game theoretic decision making, it plays a zero sum game with its internal simulation of the tracker, as described in section \ref{methodology}, and uses the outcome of this game to decide on its heading at each time step.  Fig. \ref{r_t23e3}  plots the comparative  performance of the tracker when the evader persists with this action.  The first observation that we can make from the figure is that when the evader undertakes game theoretic decision making, the tracker win percentages are the highest (compared to the  linear escape of the evader). For some speed ratios, the tracker win percentage is higher than 90\%. So game theoretic decision making seems to favour the tracker, not evader, even when both agents play their distinct zero-sum games by simulating the other.  It could also be noted that in terms of the tracker's own actions,  the tracker is better off following game theoretic decision making action for all speed ratios.  The only difference is that when the evader chooses the linear escape action, the tracker has a relatively low win percentage regardless of the action it chooses, whereas if the evader chooses game theoretic decision making action, the the tracker has a high win percentage regardless of the action it chooses. It is also interesting to note that while using game theoretic decision making favours the tracker, it gains the strongest advantage when the speed ratio is closer to one.

\begin{figure}
    \centering
    \includegraphics[scale=0.45]{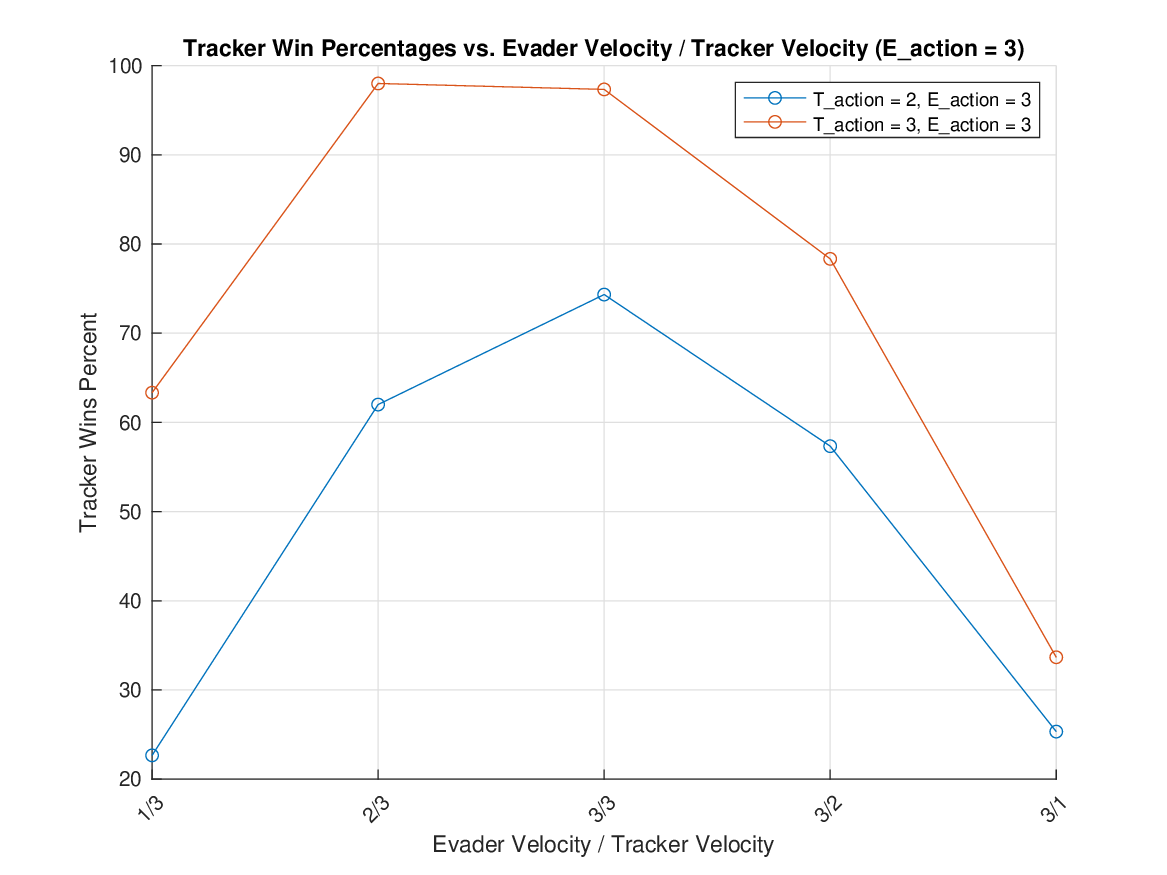}
    \caption{Comparing actions by tracker when evader chooses  game theoretic decision making. Note that for all speed ratios, the tracker is better off making decisions by using game theory.}
    \label{r_t23e3} 
\end{figure}

The summary of these two scenarios is that regardless of which action the evader takes, the tracker is usually better off undertaking game theoretic decision making at each time step compared to following a simple optimisation method based on minimising the covariance of the evader state estimate. The advantage of using game theory in decision making (for tracker) is more pronounced when the evader also uses game theoretic decision making.

\subsection{Optimal action by evader}

Here we compare different actions the evader could take if the tracker chooses a particular action. In the previous subsection, it already became clear that regardless of the tracker's action, the evader benefits most from a linear escape strategy. In this subsection we will dissect this observation further by methodically considering all scenarios faced by the evader one by one.

\subsubsection{Optimal action by evader  if tracker chooses covariance minimisation}

Fig. \ref{r_e123t2} shows the win percentages of the evader against speed ratios, when the tracker chooses the covariance minimisation action. The different actions the evader can take are compared in terms of the evader's win percentages. As expected, we first note that for most speed ratios, the evader is better off following a linear escape strategy, which gives the evader a win percentage of 70\% or above for all speed ratios. It could be noted though that  when   $v_E : v_T   =  1 : 3$,  that is, when the tracker has a very strong speed advantage, then the evader has the game theoretic decision making as the best action. It is interesting to note that when the linear escape action is rendered less attractive to the evader since it has a strong speed disadvantage (i.e, `brute force' is less likely to help) that is when a game theoretic decision making action helps the evader more. It should also be noted that for the particular number of time steps we consider ($t=400$) the evader is often able to escape when it is much slower than the tracker. If the tracker has more time to complete its mission, then the evader is less likely to escape using linear escape strategy, and would find game theoretic decision making even more useful.

\begin{figure}
    \centering
    \includegraphics[scale=0.45]{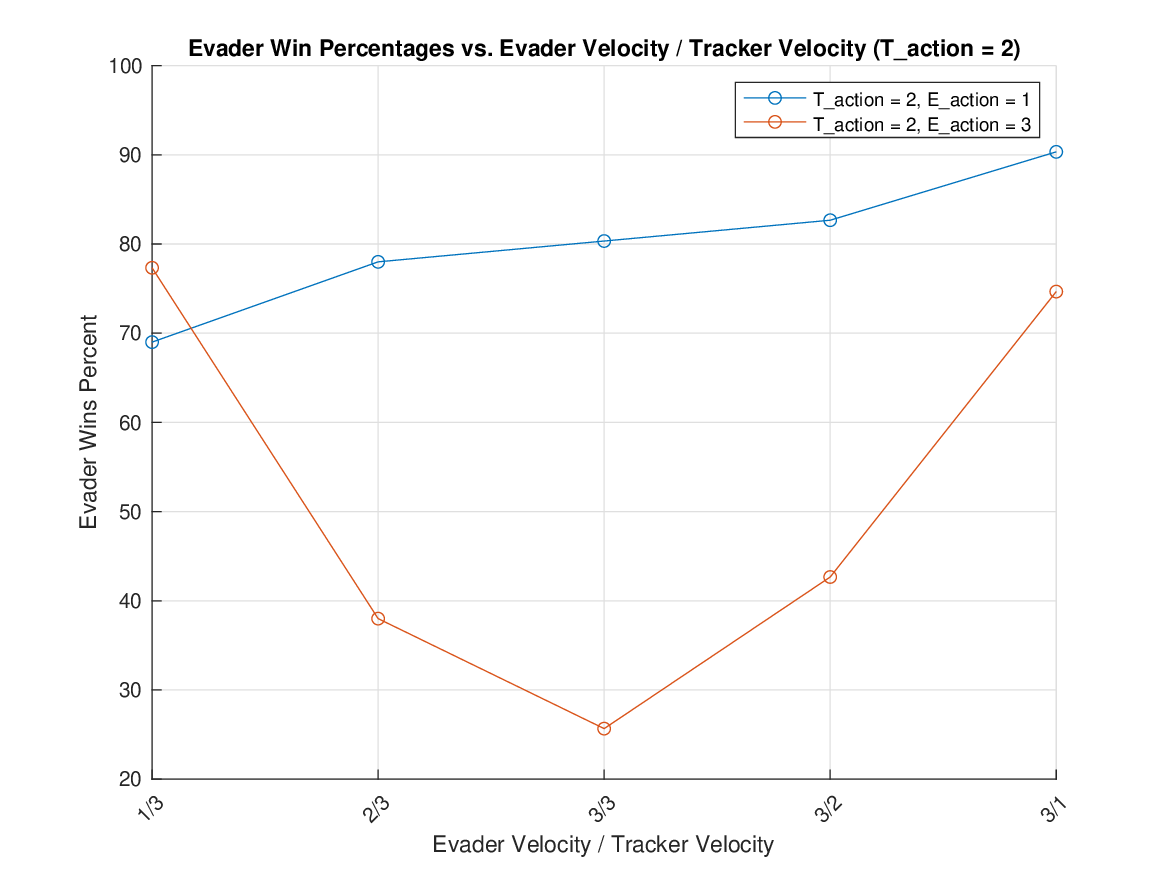}
    \caption{Comparing actions by  evader when tracker chooses covariance minimisation. It can be noted that the best action for evader is linear escape for all speed ratios, except when   it has a strong speed disadvantage, in which case it is better off undertaking game theoretic decision making.}
    \label{r_e123t2}
\end{figure}

\subsubsection{Optimal action by evader  if tracker undertakes game theoretic decision making}

Fig. \ref{r_e123t3} shows the win percentages of the evader against speed ratios, when the tracker chooses game theoretic decision making action. Again, as surmised from the previous subsection, we can indeed see that for all speed ratios, the evader is better off following a linear escape strategy, which gives the evader a win percentage of  55\% or above for all speed ratios.  When the evader undertakes game theoretic decision making, it is generally at a disadvantage compared to the tracker.

What is interesting to note is that when both parties play game theoretic decision making, the evader's win percentage drops below  10\% for speed ratios around unity. This was observed from the point of view of  the tracker in the previous subsection anyhow. However, when the speed ratio is strongly in favour of either party, the evader's win percentage improves.

We could observe that if the evader speed is three times the tracker speed, the evader can win more than 60\% of the time even when both parties undertake game theoretic decision making. This makes sense. Why, then does the evader win close to 40\% of the time when the speed ratio is    $v_E : v_T   =  1 : 3$, that is, when the evader is at a severe speed disadvantage? It appears that compared to more `predictable' scenarios, the evader is better off playing game theory when `brute force' (in this case, speed) is not in its favour. That is, the game theoretic decision making process has the effect of mitigating tremendous advantages to either party, and creating a more level playing field. Note however that when the speed ratio is in favour of the tracker, the tracker always wins more than half of the engagements, as would be expected.  However game theoretic decision making seems to somewhat negate the imbalance created by speed advantages, and seems to give a fighting chance to the party which is affected by a severe speed disadvantage.

\begin{figure}
    \centering
    \includegraphics[scale=0.45]{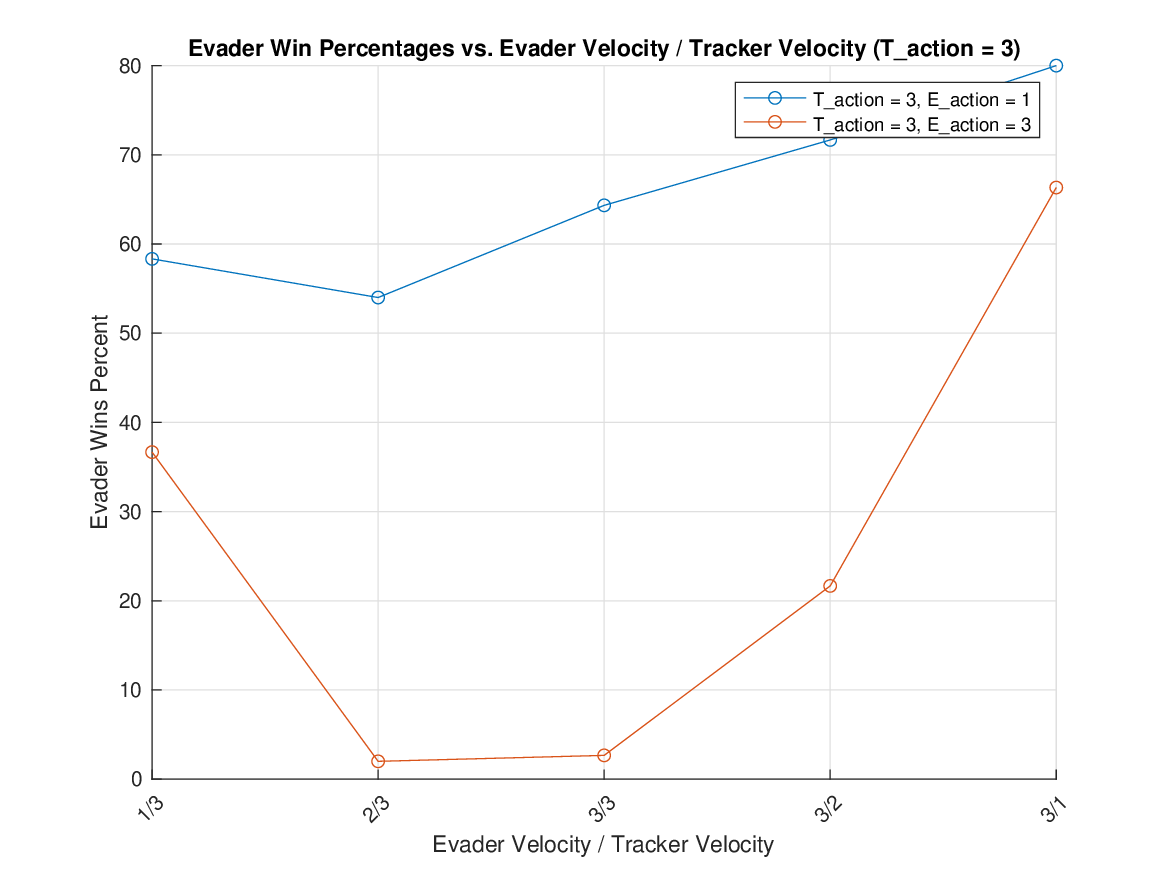}
    \caption{Comparing actions by  evader when tracker chooses game theoretic decision making. It can be noted that the best action for evader is linear escape for all speed ratios. When both tracker and player undertake game theoretic decision making, the evader loses more often, though interestingly the evader's chances of winning are better if the speed ratio is strongly favourable to either party.}
        \label{r_e123t3}
\end{figure}

The general observation that we can make here is that game theoretic decision making usually creates some opportunity to win when a player (tracker or evader) is severely disadvantaged. When non-game theoretic actions would have resulted in a player losing with near certainty, game theoretic decision making gives that player a chance to win. In terms of the tracker, game theoretic decision making worked best  for it in all scenarios. For the evader, the linear escape strategy usually worked better, given that it has a head-start and the simulation time was limited, and a `draw' was counted as a win for the evader.  So the evader just needed to avoid losing, and could try fleeing with unchanging heading. However, when the scenario was dire for the evader (for example, when it had an extreme speed disadvantage), game theoretic decision making  helped improve its chances of winning, comparatively.

\subsection{Summary of Findings} \label{summary}

The numerical results present key insights into the effectiveness of various courses of actions, in terms of decision making, that the tracker or evader can undertake.

For the tracker, the findings emphasize the importance of  action selection. The tracker is always better off undertaking game theoretic decision making compared to covariance minimisation, regardless of the decision making action the evader chooses, and regardless of the speed ratios. This is an extremely important result, particularly since a lot of tracking-related research is undertaken from the tracker's perspective. These results therefore show that the presented sequential game framework outperforms more traditional optimisation techniques which are commonly employed in tracking, and the presented framework is very valuable in improving the win percentages for the tracker, including when it is at  a severe speed disadvantage.

For the evader, the results consistently indicate that simpler tactics, such as linear escape, yield the highest win percentages for most speed ratios (only when the evader has a strong speed disadvantage does it benefit from employing game theoretic decision making). However, it is essential to acknowledge the limitations of such tactics in real-world scenarios. While linear escape consistently performs well in controlled simulations, its straightforward nature may not account for the unpredictability and adaptability required in dynamic environments. Conversely, although the game-theoretic decision making procedure results in weaker performance for the evader in general,  it demonstrates a relative improvement when the evader's speed disadvantage is severe.  This may hint at the potential benefits of adaptive strategies that could be better suited to scenarios where linear escape is insufficient due to real-world complexities. Therefore, while linear escape appears effective in these simulations, particularly with limited simulation time,  its practicality may be constrained by the need for more nuanced decision-making in unpredictable environments, or situations where the evader needs to evade the tracker for longer periods of time,  where more complex decision making procedures involving game theory might offer increased benefits in disadvantaged contexts.


\section{Conclusions}  \label{conclusions}

In this paper we presented a sequential game framework for target tracking, and compared the effectiveness of game theory-based decision making,  with simpler optimization-based decision making approaches. Taking a maritime target tracking environment as an example, we formulated a tracking problem whereby a single intelligent tracker attempts to track a single intelligent evader.  It was assumed that both the tracker and the evader employ passive sensing, and use the CKF algorithm to process the noisy bearing information obtained from their sensors. Based on this, it was further assumed that at any point in time, the tracker and the  evader have an internal simulated model of each other, and can play a zero sum game with these internal models, and use the game theoretic solutions they derive in decision making.

We modelled the tracker as being able to make decisions in two ways: 1) by following a simple optimisation procedure at each time step, whereby the tracker chooses a heading which minimises the covariance of the evader postion estimate, thus minimising the uncertainty about the evader. 2)  by playing a zero sum game with its internal simulated model of the evader at each time step, whereby the tracker chooses a heading which optimises its pay-off in this game. Similarly we modelled the evader as being able to make decisions in two ways: 1) by following a simple linear escape procedure, whereby the evader, once it becomes aware of being tracked, continues with the same speed and heading until it is caught or it escapes.  2) by playing a zero sum game with its internal simulated model of the tracker at each time step,  whereby the evader chooses a heading which optimises its payoff in this game. The idea behind simulating these decision making procedures was to compare the effectiveness of the sequential game-based decision making framework with more traditional optimisation-based decision making procedures, and to highlight under what circumstances the sequential game-based decision making framework is more effective in making decisions, for both tracker and the evader.

We simulated the above tracking problem in an infinite grid, with randomised starting positions which ensured a minimal starting distance between the tracker and the evader. We defined winning criteria for the tracker and the evader by considering three separate metrics, which were defined in terms of tracking uncertainty, tracking error, and the Euclidean distance between the tracker and evader, respectively. For each of these criteria, minimal and maximal thresholds were defined, and the particular simulation run was considered a win for the the tracker if one of the minimal thresholds were reached,  and a win for the evader if one of the maximal thresholds were reached. If no minimal or maximal threshold was reached at the end of the simulation, it was also considered a win for the evader, since the tracker was assumed to have failed in its mission. We varied the velocities of both tracker and evader, and considered a range of speed ratios between the tracker speed and the evader speed. For each particular speed ratio, we evaluated the effectiveness of various decision making procedures for both tracker and evader, and determined which decision making action (procedure) was optimal for a particular player for a particular speed ratio.

In the case of the tracker, we found that the game theoretic decision making action is superior to the covariance minimisation action in  all scenarios. Therefore it was demonstrated that the presented sequential game theoretic framework is quite effective in improving the outcomes for the tracker. On the other hand, we found that the linear escape action was the most effective for the evader in almost all scenarios. Nevertheless it was found that  game theoretic decision making could be better for the evader in scenarios where the evader has a significant speed disadvantage. It should be noted however, that given there is a limited number of time steps within which the tracker must satisfy a win criteria (this represents the time constraint in the real world within which the tracker must track the evader), and any outcome where either party fails to meet the winning criteria is considered a win for the evader by default, that it could be argued that the evader starts with an inherent advantage. Given this inherent advantage, the evader is often able to get away using a linear escape strategy. It is postulated that if the experiments are simulated for a longer time period, then the linear escape action would not be as effective.  In any case, it could be argued that the evader, too, seems to benefit more from game theoretic decision making, when it does not have a strong inherent advantage in terms of time limits or speed ratio.

Therefore in summary, it could be argued that game theoretic decision making, as presented by the sequential game framework in this paper, is particularly useful for an actor, when it has inherent disadvantages in terms of speed, starting position, or time constraints. The tracker has an inherent disadvantage because the onus is on the tractor to satisfy the winning criteria within limited time, and thus the tracker finds the game theoretic decision making framework more useful. In scenarios where the evader has also an inherent disadvantage, such as having  a lower speed compared to the tracker, the evader also finds the game theoretic decision making framework relatively more useful. The simplest interpretation of these observations is that, a complex decision making procedure such as the game theoretic decision making procedure presented here, is less useful in scenarios where it is not needed. If an agent has  an inherent advantage  such that it is more likely to win anyhow, it is more beneficial for it to use straightforward tactics, or simpler decision making procedures based on single quantity optimisation. On the other hand, when brute force will not suffice, that is, an agent is unlikely to win based on its speed or starting position alone, that is when relatively more complex decision making procedures based on the presented game theoretic framework increase its chances of winning. Where there is less brawn, there must be more brain, to improve the chances of winning!

It should be noted that the presented work has a number of limitations, in terms of conceptualisation as well as simulation. In terms of conceptualisation, the primary limitation is that we are assuming that an agent is committed to a particular way of decision making, a certain decision making action or procedure, from beginning to end. This need not be the case. A tracker or evader could use a simpler optimisation-based decision making at certain time steps, and game theoretic decision making at other time steps. Adapting such a mixed decision making approach probably is more optimal than sticking with a single decision making procedure, and is something that needs to be explored in future. In terms of simulation, even though the starting positions were randomised, they were constrained such that a certain minimal distance was maintained between the tracker and evader. It would be interesting to see if the winning percentages vary significantly if more variation in terms of starting positions is considered. Similarly, the effect of having different time windows for the tracker to succeed in its mission, and the effect of the length of this time window in the tracker's winning percentage, need to be analysed. Therefore the simulation experiments could be expanded in a number of ways. These are some of the possible avenues for the future expansion of this work.

}

\section*{Acknowledgment}

DL and MP would like to thank the Defence Science and Technology Group (DSTG)  for providing funding for this work under the STaRShot grant 10397.

\bibliographystyle{IEEEtran}
\bibliography{Tracking_game}

\end{document}